\documentclass[a4paper,12pt]{article}

\usepackage[dvips]{graphicx}

\usepackage{amsmath}

\begin{document} 

\begin{titlepage}

\hrule 
\leftline{}
\leftline{Chiba Univ. Preprint
          \hfill   \hbox{\bf CHIBA-EP-127}}
\leftline{\hfill   \hbox{hep-th/0105262}}
\leftline{\hfill   \hbox{May 2001}}
\vskip 5pt

\hrule 
\vskip 1.0cm
%%%%%%%%%
% Title %
%%%%%%%%%
\centerline{\large\bf
  Renormalizable Abelian-Projected Effective Gauge Theory}
\centerline{\large\bf
  Derived from Quantum Chromodynamics II} 

\vskip 1cm

\centerline{{\bf
Toru Shinohara${}^{1,\dagger}$
}}
\vskip 1cm
\begin{description}
\item[]{\it 
$^1$ Graduate School of Science and Technology,
  Chiba University, Chiba 263-8522, Japan
  }
\end{description}

%%%%%%%%%%%%%%%%%%%%%%%%%%%%%
\centerline{{\bf Abstract}} %
%%%%%%%%%%%%%%%%%%%%%%%%%%%%%

\vskip 0.5cm

In the previous paper\cite{KS00b}, we derived
the Abelian projected effective gauge theory
as a low energy effective theory of the $SU(N)$ Yang-Mills theory
by adopting the maximal Abelian gauge.
At that time, we have demonstrated the multiplicative renormalizability
of the propagators for the diagonal gluon
and the dual Abelian anti-symmetric tensor field.
In this paper,
we show the multiplicative renormalizability
of the Green's functions also for the off-diagonal gluon.
Moreover we complement the previous results
by calculating the anomalous dimension and
the renormalization group functions
which are undetermined in the previous paper.


\vskip 0.2cm
\hrule  
${}^\dagger$ 
  E-mail: {\tt sinohara@cuphd.nd.chiba-u.ac.jp},
          {\tt sinohara@graduate.chiba-u.jp}

\vskip 0.5cm  

\end{titlepage}

%%%%%%%%%%%%%%%%%%%%%%%%%%%%%%%%%%%%%%%%%%%%%%%%%%%%%%%%%%%%%%%%%%%%%%
%%%%%%%%%%%%%%%%%%%%%%%%%%%%%%%%%%%%%%%%%%%%%%%%%%%%%%%%%%%%%%%%%%%%%%

%\tableofcontents
%\newpage

%%%%%%%%%%%%%%%%%%%%%%%%
\section{Introduction} %
%%%%%%%%%%%%%%%%%%%%%%%%
Quark confinement and spontaneous breakdown of chiral symmetry
are very important problems in the low energy Quantum Chromodynamics (QCD).
It is difficult to analyze QCD in the low energy region
due to strong coupling constant,
while it is easy to analyze QCD in the high energy region
due to the asymptotic freedom of coupling constant.
Thus the effective theory is often used
in order to investigate the phenomena in the low energy region of QCD.
In fact,
it is well-known that the problems mentioned above
are qualitatively explained by dual Ginzburg-Landau (DGL) theory
which describes the dual superconductivity
due to the dual Meisner effects.\cite{Nambu,tHooft81,Mandelstam}
Therefore we expect
that the DGL theory is a candidate
of a low energy effective theory of QCD.
However, the analytical derivation of DGL from QCD is not achieved
so that we have tried to construct the interpolated effective theory
connecting QCD and DGL in the series of our works\cite{KS00b, KondoI}.

In the previous paper\cite{KS00b},
which is referred to as (I) hereafter,
we have derived Abelian projected effective
gauge theory (APEGT) from the $SU(N)$ Yang-Mills theory
by making use of the maximal Abelian (MA) gauge\cite{KLSW87}
based on an idea of Abelian projection\cite{tHooft81}.
The APEGT is originally derived for the $SU(2)$ gauge group
in Ref.~\cite{KondoI}
and extended for the $SU(N)$ gauge group in (I).

The original Yang-Mills theory includes
the diagonal gluon, the off-diagonal gluon,
the diagonal ghost and the off-diagonal ghost.
According to the Abelian dominance\cite{EI82},
the diagonal components play the dominant roles
in the low energy region of QCD,
while the off-diagonal components
hardly affect the phenomena in the low energy region.
The mechanism of Abelian dominance can be understood by the fact
that the off-diagonal fields become massive,
while the diagonal fields remain massless or have smaller masses.
We derived the dynamical mass generations of off-diagonal fields
in Ref.~\cite{KS00a} in agreement with numerical simulations
on a lattice\cite{AS99}.
By introducing a dual Abelian anti-symmetric tensor fields
together with two arbitrary parameters $\rho$ and $\sigma$
as a dual of the composite operator of off-diagonal gluons
and by integrating out off-diagonal gluons and ghosts,
we have obtained the APEGT which includes only the diagonal gluon and
the dual Abelian tensor field.
Moreover, by integrating out the diagonal gluon,
we have arrived at the DGL-like effective theory
as demonstrated in Refs.~\cite{KS00b, KondoI}.

In (I), we calculated the beta-function, the anomalous dimensions
of the diagonal gluon and dual Abelian tensor fields,
and the renormalization group (RG) functions of
the gauge fixing parameter for the diagonal gluon
and an arbitrary parameter $\rho$
and demonstrated the multiplicative renormalizability
for the diagonal gluon and the dual Abelian tensor field propagators.
In this paper, moreover,
we show the multiplicative renormalizability
for the Green's functions including not only the diagonal fields
but also off-diagonal gluons,
and calculate the anomalous dimension of off-diagonal gluon
and the RG functions of the gauge fixing parameter
for the off-diagonal gluon and the remaining parameter $\sigma$.
%We adopt the {\it modified} MA gauge\cite{KS00a,SIK01a}
%instead of the {\it naive} MA gauge adopted in (I).
%However, the results obtained in (I) are not changed by this modification.
The results obtained here complement the previous results.

%%%%%%%%%%%%%%%%%%%%%%%%%%%%%%%%%%%%%%%%%%
\section{Modified maximal Abelian gauge} %
%%%%%%%%%%%%%%%%%%%%%%%%%%%%%%%%%%%%%%%%%%
%In this section 
We construct the gauge fixing (GF) and the associated Faddeev-Popov (FP) ghost term for the MA gauge.
To analyze the non-Abelian gauge theory in the MA gauge,
%it is often useful to 
we distinguish the color indices as follows:
\begin{equation}
\left\{
\begin{array}{ccl}
A,B,C,\ldots & \rightarrow & SU(N), \cr
i,j,k,\ldots & \rightarrow & U(1)^{N-1}\quad(\mbox{diagonal}), \cr
a,b,c,\ldots & \rightarrow & SU(N)/U(1)^{N-1}\quad(\mbox{off-diagonal}).
\end{array}
\right.
\end{equation}

First, we define the MA gauge condition.
The MA gauge is obtained by minimizing the functional $R[A^U]$
with respect to the local gauge transformation $U(x)$ of $A_{\mu}^a$.
Here, $R[A]$ is defined as the functional of off-diagonal gluons,
\begin{equation}
R[A]:=\int d^4x\frac12A_\mu^a(x)A^{\mu a}(x).
\end{equation}
Then we obtain the differential form of the MA gauge condition,
\begin{equation}
F^a[A]:=D_\mu A^{\mu a}=0,
\end{equation}
where we defined the covariant derivative with respect to
the diagonal gluon $A_\mu^i$ as
\begin{equation}
D_\mu{\mit\Phi}^A
 :=\left(\partial_\mu\delta^{AB}
         +gf^{AiB}A_\mu^i\right)
   {\mit\Phi}^B,
\end{equation}
for an arbitrary operator ${\mit\Phi}^A$.
The MA gauge condition partially fixes the color rotational symmetry
from $SU(N)$ to $U(1)^{N-1}$.
In fact, the residual $U(1)^{N-1}$ symmetry,
\begin{equation}
\left\{
\begin{array}{l}
\delta A_\mu^i=\partial_\mu\theta^i, \cr
\delta{\mit\Phi}^i=0
  \quad(\mit\Phi^i=\phi^i,C^i,\bar C^i), \cr
\delta{\mit\Phi}^a=-gf^{abi}{\mit\Phi^b}\theta^i
  \quad(\mit\Phi^a=A_\mu^a,\phi^a,C^a,\bar C^a), \cr
\end{array}
\right.
\label{eq:U(1) transformation}
\end{equation}
is not fixed by the MA gauge fixing condition.
Here $A_\mu$, $\phi$, $C$ and $\bar C$ is gluon,
Nakanishi-Lautrup field, ghost and antighost respectively.
In order to complete the gauge fixing, we must fix the residual
$U(1)^{N-1}$ symmetry.
In this paper, we employ the additional gauge fixing condition
$F^i[A_\mu]:=\partial_\mu A^{i\mu}=0$
for fixing the residual $U(1)^{N-1}$ symmetry.

Next, we construct the GF+FP term for the MA gauge.
In (I), we adopted the {\it naive} MA gauge fixing term
\begin{equation}
S_{\rm MA}
 :=-\int d^4x
     i\mbox{\boldmath$\delta$}_{\rm B}
     \left[\bar C^a
           \left(D^\mu A_\mu^a
                 +\frac\alpha2\phi^a\right)
           +\bar C^i
            \left(\partial^\mu A_\mu^i
                  +\frac\beta2\phi^i\right)\right],
\label{eq:S_MA}
\end{equation}
where $\mbox{\boldmath$\delta$}_{\rm B}$ is the BRST transformation.
However, the GF+FP term~(\ref{eq:S_MA}) is not appropriate
from the viewpoint of multiplicative renormalizability of the theory.
The reason is as follows.
The divergent contribution
proportional to a ghost-antighost self interaction term
$(f^{iab}\bar C^aC^b)^2$
is generated as a quantum correction
due to the existence of the interaction
$f^{iad}f^{icb}\bar C^aC^bA^{\mu c}A_\mu^d$
in the MA gauge~(\ref{eq:S_MA}).
However, the GF+FP term~(\ref{eq:S_MA}) does not include such a term
at the tree level
so that the multiplicative renormalizability
cannot be maintained in the off-diagonal gluon sector.
Therefore, in order to preserve the multiplicative renormalizability,
we should modify the GF+FP term to include such a term 
at the tree level.
In this paper,
we adopt the {\it modified} MA gauge fixing term\cite{KondoII,KS00a}
\begin{equation}
S_{\rm mMA}
 :=\int d^4x
     i\left[%\textstyle
      \mbox{\boldmath$\delta$}_{\rm B}
      \bar{\mbox{\boldmath$\delta$}}_{\rm B}
      \left(\frac12A_\mu^aA_\mu^a
                 -\frac\alpha2iC^a\bar C^a\right)
           -\mbox{\boldmath$\delta$}_{\rm B}
            \left\{C^i
            \left(\partial^\mu A_\mu^i
                  +\frac\beta2\phi^i\right)\right\}\right],
\label{eq:S_mMA}
\end{equation}
where $\bar{\mbox{\boldmath$\delta$}}_{\rm B}$ is the anti-BRST transformation.
Integrating out $\phi^i$ and $\phi^a$,
we obtained the total gauge fixing term
\begin{align}
S_{\rm mMA}
 =&\int d^4x\Bigl\{
    %-\frac14F^{\mu\nu A}F_{\mu\nu}^A
    -\frac1{2\alpha}(D^\mu A_\mu^a)^2
    -\frac1{2\beta}(\partial^\mu A_\mu^i)^2
    -ig^2f^{adi}f^{cbi}\bar C^aC^bA^{\mu c}A_\mu^d
    \nonumber\\
  &+i\bar C^aD^2C^a
    +ig\bar C^aD^\mu(f^{abc}A_\mu^bC^c)
    +\frac i2g(D^\mu A_\mu^a)(f^{abc}\bar C^bC^c)
    \nonumber\\
  &+\frac\alpha8g^2f^{abe}f^{cde}\bar C^a\bar C^bC^cC^d
    +\frac\alpha4g^2f^{abi}f^{cdi}\bar C^a\bar C^bC^cC^d
    \nonumber\\
  &+\frac\alpha8g^2(f^{abc}\bar C^bC^c)^2
    +i\bar C^i\partial^2C^i
    +i\bar C^i\partial^\mu(gf^{ibc}A_\mu^bC^c)
    \Bigr\}.
\label{eq:S_GF}
\end{align}
This term includes the desired form ghost interaction term.%
\footnote{%
The GF+FP term~(\ref{eq:S_mMA}) satisfies
the charge conjugation invariance,
global shift invariance for the diagonal ghost
and global shift invariance for the diagonal antighost.
See Ref~\cite{SIK01a} for more detail.
}
Moreover, we find that no interaction term with
diagonal ghost $C^i$ exists so that a diagonal ghost does
not appear as an internal line in the perturbative loop calculation.

%%%%%%%%%%%%%%%%%%%%%%%%%%%%%%%%%%%%%%%%%%%%%%%%%%%%%%%%%%
\section{Composite operator and Abelian auxiliary field} %
%%%%%%%%%%%%%%%%%%%%%%%%%%%%%%%%%%%%%%%%%%%%%%%%%%%%%%%%%%
\label{sec:auxiliary}
Here, we consider the Green's function including
Abelian composite operator which is made of off-diagonal fields
and invariant
under the $U(1)$ gauge transformation~(\ref{eq:U(1) transformation}).
Then we introduce a dual Abelian tensor field $B_{\mu\nu}^i$
as an auxiliary field corresponding to such a composite operator.

According to the Abelian dominance
in the low energy region of QCD,
an elementally off-diagonal field hardly affect the physics alone.
However, the Abelian composite operator
which is composed of off-diagonal fields
can affect the low energy physics as well as elementary Abelian fields.
There are four candidates of the Abelian composite operator
made of off-diagonal fields with mass dimension 2, that is,
%\begin{equation}
$f^{ibc}A_\mu^bA_\nu^c$, $A_\mu^aA^{\mu a}$,
$f^{ibc}\bar C^bC^c$ and $\bar C^aC^a$.
%\end{equation}
In this paper we focus on the two-rank anti-symmetric composite operator
%\begin{equation}
$O_{\mu\nu}^i:=f^{ibc}A_\mu^bA_\nu^c$
%\label{eq:composite_O}
%\end{equation}
since this operator is very important to construct
the dual Abelian gauge theory of Yang-Mills theory.%
\footnote{%
Other three scalar composite operators are meaningful
in the different sense.
In the non-perturbative consideration
by making use of effective potential,
the dynamical mass of the off-diagonal fields are caused
through the pair condensation of the off-diagonal ghost
and antighost.
However, in this paper,
all calculations are based on the perturbative method
and the non-zero masses of the off-diagonal fields are assumed.
}
In order to consider the Green's functions with Abelian
composite operator $O^i$, we introduce a source term $S_K$ for $O^i$
as well as $S_J$ for elemental fields:
\begin{equation}
Z[J,K]=\int\!\!D{\mit\Phi}\exp\{iS_{\rm YM}+iS_{\rm mMA}+iS_J+iS_K\}.
\label{eq:partition function}
\end{equation}
Where $S_{\rm YM}:=-\frac14F_{\mu\nu}^AF^{\mu\nu A}$ is Yang-Mills action
and gauge fixing terms $S_{\rm mMA}$
have already been given in (\ref{eq:S_mMA}).
$S_J$ is a source term for the elementary fields,
%\begin{equation}
$
S_J
 :=\int d^4x\left(A_\mu^iJ^{\mu i}+A_\mu^aJ^{\mu a}\right)
$,
%\end{equation}
and $S_K$ is a source term for the composite operator:
\begin{equation}
S_K
 :=\int d^4x\frac i2{}^\ast\!O_{\mu\nu}^iK^{\mu\nu i},
\label{eq:S_K}
\end{equation}
where we define the Hodge dual
of arbitrary two-rank anti-symmetric tensor as
$
%\begin{equation}
{}^\ast\!O_{\mu\nu}
 :=\frac12\epsilon_{\mu\nu\rho\sigma}O^{\rho\sigma}
%\end{equation}
$.
However, we notice that the composite source term~(\ref{eq:S_K})
is insufficient from the viewpoint of multiplicative renormalizability.
Indeed, such a source term~(\ref{eq:S_K}) generate
the divergent contributions not only proportional to
$O_{\mu\nu}^iK^{\mu\nu i}$ but also proportional to
$\partial_\mu A_\nu^i K^{\mu\nu i}$ and $K_{\mu\nu}^iK^{\mu\nu i}$.
%must be generated.
(See Fig.~\ref{fig:composite source})
\unitlength=0.001in
%%%%%%% BEGIN FIGURE (Feynman Rules) %%%%%%%%%%
\begin{figure}[tb]
\begin{center}
\begin{picture}(4500,800)%(0,-3000)
%\put(0,800){\tframe[500][100](4500,800)}%
%-----------%
% Graph (a) %
%-----------%
%\put(0,600){\mbox{\large(a)}}%
\put(350,0){%
%   \put(0,0){\epsfysize=5mm \epsfbox{apropa.eps}}%
   \put(0,0){\includegraphics[height=20mm]{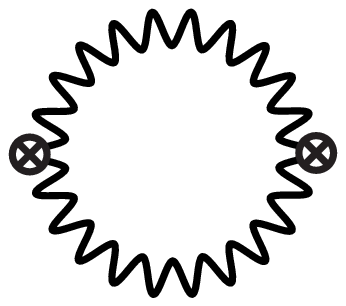}}%
   }%
%-----------%
% Graph (b) %
%-----------%
%\put(1450,600){\mbox{\large(b)}}%
\put(1700,0){%
   \put(0,0){\includegraphics[height=20mm]{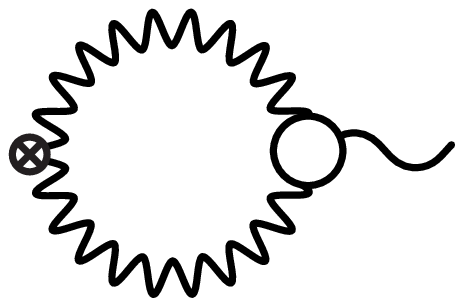}}%
   }%
%-----------%
% Graph (c) %
%-----------%
%\put(3000,600){\mbox{\large(c)}}%
\put(3250,0){%
%   \put(0,0){\epsfysize=2mm \epsfbox{ghost.eps}}%
   \put(0,0){\includegraphics[height=20mm]{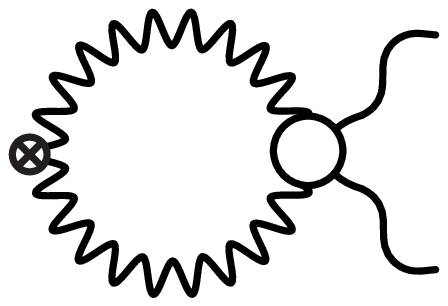}}%
   }%
\end{picture}
\caption[]{The wavy line denotes the gluon
and the circled cross denotes the composite source $K_{\mu\nu}^i$.}
\label{fig:composite source}
\end{center}
\end{figure}
%%%%%%% END FIGURE (Feynman Rules) %%%%%%%%%%
Therefore, in order to maintain the multiplicative renormalizability,
we should re-define the composite source term as
\begin{equation}
S_K
 :=\int d^4x\left(\frac i2{}^\ast\!Q_{\mu\nu}^iK^{\mu\nu i}
                  +\frac14K_{\mu\nu}^iK^{\mu\nu i}\right),
\end{equation}
where we define new composite operator
\begin{equation}
Q_{\mu\nu}^i:=\rho f_{\mu\nu}^i+\sigma f^{ibc}A_\mu^bA_\nu^c,
\end{equation}
with new two parameters $\rho$ and $\sigma$.%
\footnote{Three divergent contributions
          shown in Fig.~\ref{fig:composite source} can be absorbed
          by renormalizing the composite source $K_{\mu\nu}^i$
          and two new parameters $\rho$ and $\sigma$.}
Then we can obtain the connected Green's functions including the
Abelian composite operator $Q_{\mu\nu}^i$ by
differentiating $W:=-i\ln Z$ with respect to $K_{\mu\nu}^i$.

Now we try to define the effective action.
The existence of the quadratic source term such as the last term of $S_K$
prevents us from performing the Legendre transformation.
Therefore we introduce a dual Abelian tensor field as an auxiliary field
in order to put out quadratic term of $K_{\mu\nu}^i$.
By making use of the identity:
\begin{equation}
{\cal N}
\int DB_{\mu\nu}^i
 \exp i\int d^4x
 \left\{-\frac14
  \left(B_{\mu\nu}^i
        -{}^\ast\!Q_{\mu\nu}^i
        -K_{\mu\nu}^i
  \right)^2
 \right\}
\equiv1,
\end{equation}
with an appropriate constant ${\cal N}$,
we obtain
\begin{align}
Z[J,K]
 &={\cal N}
   \int\!\!D{\mit\Phi}D\!B_{\mu\nu}^i
    \exp\{iS_{\rm inv}+iS_{\rm GF}+iS_J
          +i\int d^4x\frac12B_{\mu\nu}^iK^{\mu\nu i}\},
%   \nonumber\\
% &=\int\!\!D{\mit\Phi}D\!B_{\mu\nu}^i
%    \exp\{iS_{\rm inv}+iS_{\rm GF}+iS_J
%          +i\int d^4x{}^\ast\!B_{\mu\nu}^i{}^\ast\!K^{\mu\nu i}\}.
\label{eq:partition function with B}
\end{align}
where we defined
\begin{align}
S_{\rm inv}
 =\,&\int d^4x\Bigl\{
    -\frac14\left[D_\mu A_\nu^a
                  -D_\nu A_\mu^a
                  +gf^{abc}A_\mu^bA_\nu^c\right]^2
    \nonumber\\
 & -\frac{1-\rho^2}4\left(f_{\mu\nu}^i\right)^2
    -\frac{1-\rho\sigma}2gf_{\mu\nu}^if^{ibc}A^{\mu b}A^{\nu c}
    -\frac{1-\sigma^2}4g^2\left(f^{ibc}A_\mu^bA_\nu^c\right)^2
    \nonumber\\
 & -\frac14\left(B_{\mu\nu}^i\right)^2
    -\frac14\epsilon^{\mu\nu\rho\sigma}B_{\mu\nu}^i
     \left[\rho f_{\mu\nu}^i+\sigma gf^{ibc}A_\mu^bA_\nu^c\right]
     \Bigr\},
\label{eq:S_inv}
\end{align}
with the Abelian field strength
$
%\begin{equation}
f_{\mu\nu}^i
 :=\partial_\mu A_\nu^i-\partial_\nu A_\mu^i%,
%\end{equation}
$.
It is easily confirmed that we can reproduce the original
partition function~(\ref{eq:partition function})
by integrating out $B_{\mu\nu}^i$ in~(\ref{eq:partition function with B}).
%and taking a limit of $K_{\mu\nu}^i\rightarrow0$.

Now we can perform the Legendre transformation as
\begin{equation}
{\mit\Gamma}[\bar A_\mu^i,\bar A_\mu^a,\bar B_{\mu\nu}^i]
 :=W[J,K]
   %-i\ln Z[J,K]
   -\int d^4x\left(\bar A_\mu^iJ^{\mu i}
                   +\bar A_\mu^aJ^{\mu a}
                   +\frac12\bar B_{\mu\nu}^iK^{\mu\nu i}\right),
\end{equation}
where we defined the background fields:
\begin{equation}
\bar A_\mu^A:=\frac{\delta}{\delta J^{\mu A}}W[J,K],
\quad
\bar B_{\mu\nu}^i:=\frac{\delta}{\delta K^{\mu\nu i}}W[J,K].
\end{equation}
We obtain one particle irreducible (1PI) Green's function
by making use of the effective action
${\mit\Gamma}[\bar A_\mu^i,\bar A_\mu^a,\bar B_{\mu\nu}^i]$.

%%%%%%%%%%%%%%%%%%%%%%%%%%%%%%%%%%%%%%%%%%%
\section{Renormalization and RG functions} %
%%%%%%%%%%%%%%%%%%%%%%%%%%%%%%%%%%%%%%%%%%%
%In this section, 
We introduce the multiplicative renormalization factors
of the fields, parameters and coupling constant in order to absorb
undesirable divergent contributions in the 1PI Green's functions.
Especially, we pay attention to the 1PI Green's functions
including the dual Abelian tensor field $B_{\mu\nu}^i$
and determine the RG function of $\sigma$
which is unfixed in (I).

%============================================================%
%\subsection{Multiplicative renormalization and counterterms} %
%============================================================%

In order to show the multiplicative renormalizability
and determine the beta function, the anomalous dimensions
and the RG functions we define the renormalized fields and parameters as
\begin{equation}
\begin{array}{c}
\bar A_\mu^i=Z_a^{1/2}\bar A_{{\rm R}\mu}^i,
\quad
\bar A_\mu^a=Z_A^{1/2}\bar A_{{\rm R}\mu}^a,
\quad
\bar B_{\mu\nu}^i=Z_B^{1/2}\bar B_{{\rm R}\mu\nu}^i,
\quad
g=Z_gg_{\rm R}, \\[4mm]
%\quad
\alpha=Z_\alpha\alpha_{\rm R},
\quad
\beta=Z_{\beta}\beta_{\rm R},
\quad
\rho=Z_\rho\rho_{\rm R},
\quad
\sigma=Z_\sigma\sigma_{\rm R}.
\end{array}
\label{eq:renormalization}
\end{equation}
Here, at the one-loop level calculations,
there is no need to take into account renormalization of each
fluctuation field explicitly.
Moreover the renormalization of the background ghost and antighost
are also irrelevant because we consider
the Green's functions including neither ghost nor antighost
in this paper.
We can determine each renormalization constant defined here
to absorb all divergent contributions calculated
in the 1PI Green's function.

\begin{figure}[tb]
\begin{center}
\begin{picture}(5300,600)(0,200)
%\put(0,600){\tframe[500][100](5300,600)}%
%===========%
% Graph (A) %
%===========%
\put(0,500){\mbox{(A)}}%
\put(0,0){%
   \put(0,0){\includegraphics[height=13mm]{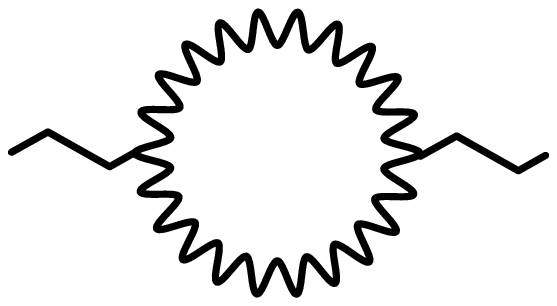}}%
   \put(50,120){\mbox{$i$}}%
   \put(850,270){\mbox{$i$}}%
   }%
%===========%
% Graph (B) %
%===========%
\put(1150,500){\mbox{(B)}}%
\put(1150,0){%
   \put(0,0){\includegraphics[height=13mm]{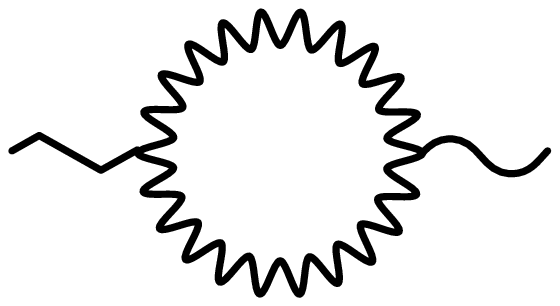}}%
   \put(50,120){\mbox{$i$}}%
   \put(850,270){\mbox{$i$}}%
   }%
%-----------%
% Graph (C1) %
%-----------%
\put(2300,500){\mbox{(C1)}}%
\put(2300,0){%
   \put(0,0){\includegraphics[height=12.5mm]{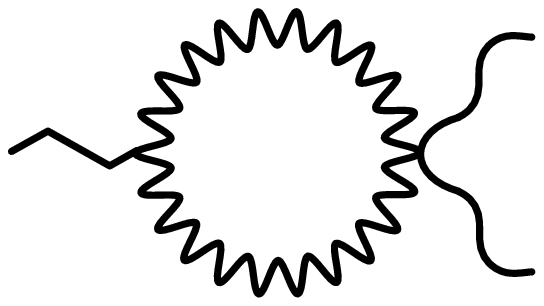}}%
   \put(50,120){\mbox{$i$}}%
   }%
%-----------%
% Graph (C2) %
%-----------%
\put(3300,500){\mbox{(C2)}}%
\put(3300,0){%
   \put(0,0){\includegraphics[height=12.5mm]{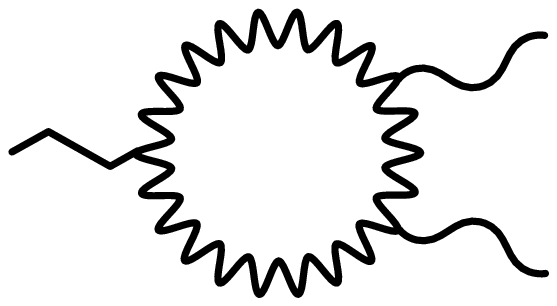}}%
   \put(50,120){\mbox{$i$}}%
   }%
%-----------%
% Graph (C3) %
%-----------%
\put(4300,500){\mbox{(C3)}}%
\put(4300,0){%
   \put(0,0){\includegraphics[height=12.5mm]{baa2.eps}}%
   \put(50,120){\mbox{$i$}}%
   \put(720,190){\mbox{$i$}}%
   }%
\end{picture}
\end{center}
\caption[]{The graphical representation of the Green's function
           including $B_{\mu\nu}^i$ at the one-loop level.
           The zig-zag line denotes the dual Abelian tensor field.
           The lines labeled ``$i$'' correspond to the diagonal fields.
           }
\label{fig:Green's functions}
\end{figure}

%====================================%
%\subsection{Green's function with $B$} %
%====================================%
Now, we consider the Green's functions including $B_{\mu\nu}^i$.
There are three divergent Green's functions
$\left<\bar B_{\mu\nu}^i\bar B_{\xi\eta}^j\right>$,
$\left<\bar B_{\mu\nu}^i\bar A_\lambda^j\right>$ and
$\left<\bar B_{\mu\nu}^i\bar A_\xi^a\bar A_\eta^b\right>$.
The graphical representation is given
in Fig.~\ref{fig:Green's functions}.
%%%%%%% BEGIN FIGURE (Vacuum Polarization) %%%%%%%%%%
%%%%%%% END FIGURE (Vacuum Polarization) %%%%%%%%%%
The graphs (A), (B) and (C)s correspond to the first, second
and third graphs in Fig.~\ref{fig:composite source} respectively.

We have already calculated the vacuum polarization graphs (A) and (B)
in (I)
together with the vacuum polarization for the diagonal gluon propagator
$\left<A_\mu^iA_\nu^j\right>$.
And we determined the renormalization constants
$Z_a$, $Z_B$, $Z_g$ and $Z_\beta$ there.
It is remarkable
that the results in (I) paper are also valid here
despite that we adopt the {\it modified} MA gauge
in this paper which differs
from the {\it naive} MA gauge in (I).
Therefore the remaining renormalization constants
to be determined in this paper
are $Z_A$, $Z_\alpha$ and $Z_\sigma$.
First, $Z_A$ and $Z_\alpha$ can be determined by requiring that
the transverse and longitudinal mode of the off-diagonal gluon propagator
are convergent.
Then we can determine $Z_\sigma$ by requiring the Green's function
$\left<\bar B_{\mu\nu}^i\bar A_\xi^a\bar A_\eta^b\right>$
is convergent.

\begin{figure}[tb]
\begin{center}
\begin{picture}(5800,600)(0,200)
%\begin{picture}(5800,400)(0,200)
%\put(0,600){\tframe[500][100](5800,600)}%
%-----------%
% Graph (D1) %
%-----------%
\put(0,500){\mbox{(D1)}}%
\put(0,0){%
   \put(0,0){\includegraphics[height=12mm]{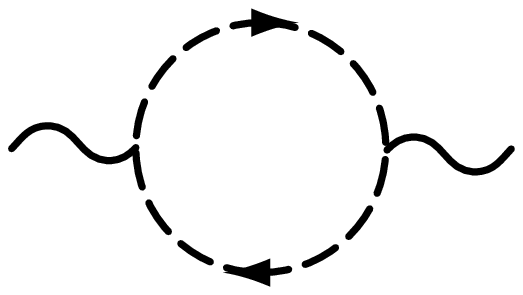}}%
   }%
%-----------%
% Graph (D2) %
%-----------%
\put(950,500){\mbox{(D2)}}%
\put(950,0){%
   \put(0,0){\includegraphics[height=12.5mm]{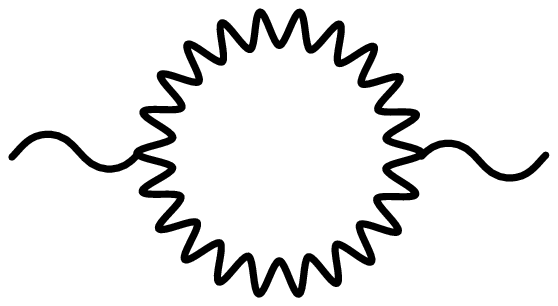}}%
   }%
%-----------%
% Graph (D3) %
%-----------%
\put(1950,500){\mbox{(D3)}}%
\put(1950,0){%
   \put(0,0){\includegraphics[height=12.5mm]{aa2.eps}}%
   \put(420,300){\mbox{$i$}}%
   }%
%-----------%
% Graph (D4) %
%-----------%
\put(2950,500){\mbox{(D4)}}%
\put(3050,0){%
   \put(0,0){\includegraphics[height=13mm]{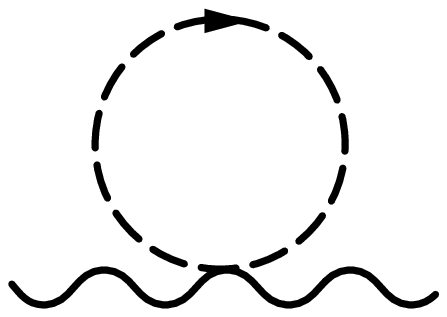}}%
   }%
%-----------%
% Graph (D5) %
%-----------%
\put(3900,500){\mbox{(D5)}}%
\put(4050,0){%
   \put(0,0){\includegraphics[height=13.5mm]{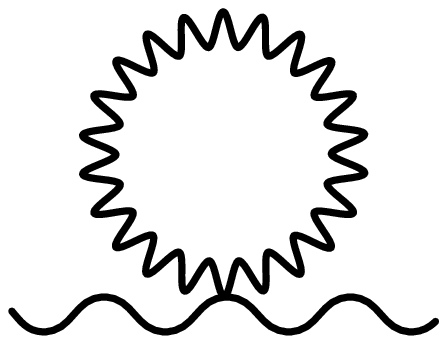}}%
   }%
%-----------%
% Graph (D6) %
%-----------%
\put(4900,500){\mbox{(D6)}}%
\put(5050,0){%
   \put(0,0){\includegraphics[height=13.5mm]{aa4.eps}}%
   \put(320,330){\mbox{$i$}}%
   }%
\end{picture}
\end{center}
\caption[]{Vacuum polarization graphs
           with respect to the off-diagonal gluon propagator
           at one-loop level.
           }
\label{fig:vacuum polarization}
\end{figure}

The divergent vacuum polarization graphs
for the off-diagonal gluons are drawn
in Fig.~\ref{fig:vacuum polarization} and
the divergent contributions to Green's function
$\left<\bar B_{\mu\nu}^i\bar A_\xi^a\bar A_\eta^b\right>$
have already been shown in Fig.~\ref{fig:Green's functions}.
%\unitlength=0.001in
%%%%%%% BEGIN FIGURE (Vacuum Polarization) %%%%%%%%%%
%%%%%%% END FIGURE (Vacuum Polarization) %%%%%%%%%%

After straightforward calculations
by making use of the dimensional regularization,
we obtain
\begin{align}
Z_A
 &=1+\frac{(g_{\rm R}\mu^{-\epsilon})^2}{(4\pi)^2\epsilon}
     \left[\frac{17}6-\frac{\alpha_{\rm R}}2-\beta_{\rm R}
          +\left(\frac53+\frac{1-\alpha_{\rm R}}2\right)(C_2(G)-2)
     \right]
    +\cdots,
\\
Z_\alpha
 &=1+\frac{(g_{\rm R}\mu^{-\epsilon})^2}{(4\pi)^2\epsilon}
     \left[\frac43-\alpha_{\rm R}-\frac3{\alpha_{\rm R}}
           +\left(\frac53+\frac{1-\alpha_{\rm R}}2
                  -\frac34\alpha_{\rm R}\right)(C_2(G)-2)
     \right]
   +\cdots,
\label{eq:Z_alpha}
   \\
Z_\sigma
 &=1+\frac{(g_{\rm R}\mu^{-\epsilon})^2}{(4\pi)^2\epsilon}
     \Bigl[\frac43
             +\frac{1+\alpha_{\rm R}}2\sigma_{\rm R}^2
             +\frac54\alpha_{\rm R}
             -\frac1{\alpha_{\rm R}}
             +\frac{\beta_{\rm R}}4
             -\frac{\beta_{\rm R}}{2\alpha_{\rm R}}
   \nonumber\\
 &\qquad\qquad\qquad\qquad
           +\left(\frac5{12}
                  +\frac{\alpha_{\rm R}}2
                  +\frac{1+\alpha_{\rm R}}4\sigma_{\rm R}^2
                  \right)(C_2(G)-2)
  \Bigr]
  +\cdots,
\end{align}
where $\epsilon:=(4-d)/2$.

%=========================%
%\subsection{RG functions} %
%=========================%
Defining the anomalous dimension $\gamma_{\mit\Phi}$ 
for each field ${\mit\Phi}$
and the RG function $\gamma_\chi$ for each parameter $\chi$ as
\begin{equation}
\gamma_{\mit\Phi}
 :=\frac12\mu\frac\partial{\partial\mu}\ln Z_{\mit\Phi},
\quad
\gamma_\chi
 :=\mu\frac{\partial\chi_{\rm R}}{\partial\mu}
  =-\chi_{\rm R}\mu\frac\partial{\partial\mu}\ln Z_{\chi},
\end{equation}
we obtain the following results.
\begin{align}
\beta(g_{\rm R})
 &:=\mu\frac{\partial g_{\rm R}}{\partial\mu}
   =-g_{\rm R}\mu\frac\partial{\partial\mu}\ln Z_g
   =-\frac{11}3C_2(G)\frac{g_{\rm R}^3}{(4\pi)^2},
\\%    \nonumber\\
\gamma_a(g_{\rm R})
 &=\frac12\mu\frac\partial{\partial\mu}\ln Z_a
  =-\frac{11}3C_2(G)\frac{g_{\rm R}^2}{(4\pi)^2},
\\%   \nonumber\\
\gamma_B(g_{\rm R})
 &=\frac12\mu\frac\partial{\partial\mu}\ln Z_B
  =\frac{1+\alpha_{\rm R}}2\sigma_{\rm R}^2
    C_2(G)\frac{g_{\rm R}^2}{(4\pi)^2},
\\%   \nonumber\\
\gamma_\rho(g_{\rm R})
 &=-\rho_{\rm R}\mu\frac\partial{\partial\mu}\ln Z_\rho
    \nonumber\\
 &=-2\rho_{\rm R}
     \left[\frac{11}6+\frac{\sigma_{\rm R}^2}2
           -2\frac{\sigma_{\rm R}}{\rho_{\rm R}}
           +\frac{1-\alpha_{\rm R}}2
            \left(\frac{\sigma_{\rm R}}{\rho_{\rm R}}
                  -\frac{\sigma_{\rm R}^2}2\right)\right]
     C_2(G)\frac{g_{\rm R}^2}{(4\pi)^2},
\\%    \nonumber\\
\gamma_{\beta}(g_{\rm R})
 &=-\beta_{\rm R}\mu\frac\partial{\partial\mu}\ln Z_A 
  =-2\gamma_a(g)\beta_{\rm R}.
\end{align}
%Thus, at one-loop level, the anomalous dimensions of off-diagonal gluons
%and two parameters $\alpha$ and $\sigma$ are given by
\begin{align}
\gamma_A
 &=-\frac{g_{\rm R}^2}{(4\pi)^2}
    \left[\frac{17}6-\frac{\alpha_{\rm R}}2-\beta_{\rm R}
          +\left(\frac53+\frac{1-\alpha_{\rm R}}2\right)(C_2(G)-2)
    \right],
    \\
\gamma_\alpha
 &=\frac{2g_{\rm R}^2}{(4\pi)^2}
   \left[\frac43\alpha_{\rm R}
         -\alpha_{\rm R}^2
         -3
         +\alpha_{\rm R}
          \left(\frac53+\frac{1-\alpha_{\rm R}}2
                -\frac34\alpha_{\rm R}\right)(C_2(G)-2)
   \right],
   \\
\gamma_\sigma
 &=\frac{2g_{\rm R}^2\sigma_{\rm R}}{(4\pi)^2}
   \biggl[\frac43
          +\frac{1+\alpha_{\rm R}}2\sigma_{\rm R}^2
          +\frac54\alpha_{\rm R}
          -\frac1{\alpha_{\rm R}}
          +\frac{\beta_{\rm R}}4
          -\frac{\beta_{\rm R}}{2\alpha_{\rm R}}
   \nonumber\\
 &\qquad\qquad\quad
          +\left(\frac5{12}
                 +\frac{\alpha_{\rm R}}2
                 +\frac{1+\alpha_{\rm R}}4\sigma_{\rm R}^2
                 \right)(C_2(G)-2)
  \biggr].
\end{align}

%=======================%
%\subsection{Discussion} %
%=======================%
All anomalous dimensions and RG functions except for $\gamma_B$,
$\gamma_\rho$ and $\gamma_\sigma$ are completely independent of
two parameters $\rho$ and $\sigma$ which were introduced
in this paper together with the Abelian auxiliary field $B_{\mu\nu}^i$.
Therefore the behavior of $A_\mu^i$, $A_\nu^a$, $g$, $\alpha$ and 
$\beta$ are not affected by introduction of such an auxiliary field.
Especially, beta function obtained here is exactly identical
to that in the ordinary Lorentz gauge in which
the $SU(N)$ color rotational symmetry is unbroken.
To the contrary, $\gamma_B$, $\gamma_\rho$ and $\gamma_\sigma$ 
depend on two parameter $\rho$ and $\sigma$.
% complicatedly.

In $SU(2)$ case, we can put Casimir operator $C_2(G)$ identical to 2.
Then RG functions of each parameter are rewritten as
\begin{align}
\gamma_{\beta}
 &=\frac{44}3\beta_{\rm R}\frac{g_{\rm R}^2}{(4\pi)^2},
\\
\gamma_\alpha
 &=-2\left[\left(\alpha_{\rm R}-\frac23\right)^2
          +\frac{23}9\right]
  \frac{g_{\rm R}^2}{(4\pi)^2},
\\%    \nonumber\\
\gamma_\rho
 &=-4\rho_{\rm R}
     \left[\frac{11}6+\frac{\sigma_{\rm R}^2}2
           -2\frac{\sigma_{\rm R}}{\rho_{\rm R}}
           +\frac{1-\alpha_{\rm R}}2
            \left(\frac{\sigma_{\rm R}}{\rho_{\rm R}}
                  -\frac{\sigma_{\rm R}^2}2\right)\right]
     \frac{g_{\rm R}^2}{(4\pi)^2},
\\
\gamma_\sigma
 &=2\sigma_{\rm R}
   \biggl[\frac43
          +\frac{1+\alpha_{\rm R}}2\sigma_{\rm R}^2
          +\frac54\alpha_{\rm R}
          -\frac1{\alpha_{\rm R}}
          +\frac{\beta_{\rm R}}4
          -\frac{\beta_{\rm R}}{2\alpha_{\rm R}}
  \biggr]
  \frac{g_{\rm R}^2}{(4\pi)^2}.
\end{align}
Firstly, we consider the gauge fixing parameters $\beta$ and $\alpha$.
The RG flow of the Abelian gauge fixing parameter $\beta$ has
a fixed point at $\beta=0$.
It is similar to Landau gauge in the ordinary Lorentz gauge.
To the contrary, the RG flow of the parameter $\alpha$ does not have
any fixed point because $\gamma_\alpha$ is always negative
for arbitrary real value of $\alpha$.
The value of $\alpha$
%grows up
increases toward the infrared region.
Therefore we cannot analyze a special point of $\alpha$
differently from the case of $\beta$,
so that we set $\beta=0$ and leave $\alpha$ unfixed
in the following discussion.

Next, we consider a parameter $\rho$.
The RG flow of $\rho$ does not have a fixed point
at $\rho=0$ unless $\sigma=0$.
Of course, before the perturbative calculations,
we can expect that such a trivial fixed point $\rho=\sigma=0$ exists
since the case of $\rho=\sigma=0$ corresponds to the case in which
Abelian auxiliary field has never been introduced.
%Moreover, an introduction of $\sigma$-term must generate
%$\rho$-term without fail because any fixed point of $\rho=0$ does not
%exist when $\sigma\ne0$.
Because of the fact that a fixed point of $\rho=0$ does not
exist when $\sigma\ne0$,
we understand that an introduction of $\sigma$-term must generate
$\rho$-term.
It is compatible with the argument in Sec.~\ref{sec:auxiliary}.

Finally, we consider a remaining parameter $\sigma$.
The RG flow of $\sigma$ has a trivial fixed point at $\sigma=0$.
However, we are not interested in such a trivial fixed point.
At $\beta=0$, we obtain
\begin{equation}
\gamma_\sigma
 =\sigma_{\rm R}
  (1+\alpha_{\rm R})
  \biggl[\sigma_{\rm R}^2
         +\frac52
          \frac{(\alpha_{\rm R}-\alpha_1)
                (\alpha_{\rm R}-\alpha_2)}%
               {(1+\alpha_{\rm R})\alpha_{\rm R}}
%          \frac{1}{(1+\alpha_{\rm R})\alpha_{\rm R}}
%          \left(
%          \frac52\alpha_{\rm R}^2
%          +\frac83\alpha_{\rm R}
%          -2
%         \right)
  \biggr]
  \frac{g_{\rm R}^2}{(4\pi)^2},
\end{equation}
where $\alpha_1=(-8-2\sqrt{61})/15$ and $\alpha_2=(-8+2\sqrt{61})/15$.
$\gamma_\sigma$ can be equal to 0 at
\begin{equation}
\sigma
 \equiv
 \sqrt{-\frac52
        \frac{(\alpha_{\rm R}-\alpha_1)
              (\alpha_{\rm R}-\alpha_2)}%
             {(1+\alpha_{\rm R})\alpha_{\rm R}}},
\end{equation}
when $\alpha_{\rm 1}<\alpha_{\rm R}<-1$ or
$0<\alpha_{\rm R}<\alpha_{\rm 2}$.
However this is not a true fixed point since $\alpha$ is always
unfixed in the MA gauge as mentioned above.

%%%%%%%%%%%%%%%%%%%%%%
\section{Conclusion} %
%%%%%%%%%%%%%%%%%%%%%%
In (I) and in this paper,
we have considered the possibility of the existence of
the Abelian composite operator composed of off-diagonal gluons
in the MA gauge.
We have introduced the dual Abelian anti-symmetric tensor field
$B_{\mu\nu}^i$
into the $SU(N)$ Yang-Mills theory as an auxiliary field
together with two new parameters $\rho$ and $\sigma$.

In (I), we demonstrated the multiplicative renormalizability
of propagators for the diagonal gluon
and the dual Abelian anti-symmetric tensor field.
At that time, we calculated the beta-function,
the anomalous dimensions
of the diagonal gluon and dual Abelian anti-symmetric tensor field,
and RG functions of the gauge parameter for diagonal gluon $\beta$
and $\rho$.
The results obtained in (I) are still valid here.

In addition to the demonstration in (I), in this paper,
we have shown the multiplicative renormalizability
for the Green's functions including not only diagonal fields
but also off-diagonal gluons.
We have calculated the anomalous dimension of the off-diagonal gluon
and the RG functions of the gauge parameter for the off-diagonal gluon
$\alpha$ and parameter $\sigma$.

From the results obtained in this paper,
we have found the following facts.
First, the RG flow of the gauge fixing parameter $\alpha$
for off-diagonal gluon does not have a fixed point anywhere.
Next, an introduction of $\sigma$ term must generate the $\rho$-term
so that we should introduce $\rho$-term at the same time
from the viewpoint of multiplicative renormalizability.
Finally, the RG flow of a parameter $\sigma$ has
non-trivial fixed points depending on $\alpha$.
The non-zero values of $\rho$ and $\sigma$ mean that
the composite operator of off-diagonal gluons plays dominant role
in the low energy region.

The results in this paper are based on the perturbative calculations
so that it is valid in the ultraviolet region.
However, these are important for the analysis in the infrared region
in order to construct the low energy effective theory
which is compatible to the original QCD.

%%%%%%%%%%%%%%%%%%%%%%%%%%%%%%%%%%%%%%%%%%%%%%%%%%%%%%%%%%%%%%%%%%%%%%
%%%%%%%%%%%%%%%%%%%%%%%%%%%%%%%%%%%%%%%%%%%%%%%%%%%%%%%%%%%%%%%%%%%%%%
%%%%%%%%%%%%%%%%%%%%%%%%%%%%%%%%%%%%%%%%%%%%%%%%%%%%%%%%%%%%%%%%%%%%%%
%%%%%%%%%%%%%%%%%%%%%%%%%%%%%%%%%%%%%%%%%%%%%%%%%%%%%%%%%%%%%%%%%%%%%%
%%%%%%%%%%%%%%%%%%%%%%%%%%%%%%%%%%%%%%%%%%%%%%%%%%%%%%%%%%%%%%%%%%%%%%

%%%%%%%%%%%%%%%%%%%%%%%%%%%%%
\section*{Acknowledgements} %
%%%%%%%%%%%%%%%%%%%%%%%%%%%%%
The author would like to thank Dr.~Kei-Ichi Kondo for useful discussions
and helpful advices.

\end{document}